\documentclass[conference]{IEEEtran}
\usepackage{amsmath,amssymb,amsfonts}
\usepackage{algorithmic}
\usepackage[ruled,vlined,linesnumbered]{algorithm2e}
\usepackage{graphicx}
\usepackage{textcomp}
\usepackage{xcolor}
\usepackage{xspace}
\usepackage{makecell}
\usepackage{balance}
\usepackage[breaklinks]{hyperref}
\usepackage{url}
\usepackage{breakurl}
\usepackage{pifont}

\newtheorem{definition}{Definition}

\begin{document}

\newcommand{\bheading}[1]{{\vspace{4pt}\noindent{\textbf{#1}}}}
\newcommand{\iheading}[1]{{\vspace{4pt}\noindent{\textit{#1}}}}

\newcommand{\etal}{\emph{et al.}\xspace}
\newcommand{\etc}{\emph{etc}\xspace}
\newcommand{\ie}{\emph{i.e.}\xspace}
\newcommand{\eg}{\emph{e.g.}\xspace}

\newcommand{\figurewidth}{\columnwidth}
\newcommand{\secref}[1]{\mbox{Sec.~\ref{#1}}\xspace}
\newcommand{\secrefs}[2]{\mbox{Sec.~\ref{#1}--\ref{#2}}\xspace}
\newcommand{\figref}[1]{\mbox{Fig.~\ref{#1}}}
\newcommand{\tabref}[1]{\mbox{Table~\ref{#1}}}
\newcommand{\lstref}[1]{\mbox{Listing~\ref{#1}}}
\newcommand{\appref}[1]{\mbox{Appendix~\ref{#1}}}
\newcommand{\algref}[1]{\mbox{Algorithm~\ref{#1}}}
\newcommand{\ignore}[1]{}

\newcommand\st[1]{\textbf{Step \ding{#1}}}

\newcommand{\flushreload}{\textsc{Flush-Reload}\xspace}
\newcommand{\Flush}{\textsc{Flush}\xspace}
\newcommand{\Reload}{\textsc{Reload}\xspace}
\newcommand{\primeprobe}{\textsc{Prime-Probe}\xspace}
\newcommand{\Prime}{\textsc{Prime}\xspace}
\newcommand{\Probe}{\textsc{Probe}\xspace}
\newcommand{\evictreload}{\textsc{Evict-Reload}\xspace}
\newcommand{\flushflush}{\textsc{Flush-Flush}\xspace}
\newcommand{\flushfunc}{\texttt{clearcache}\xspace}

\newcommand{\ias}{\text{Intel's Attestation Service}\xspace}

\newcommand{\HT}{{Hyper-Threading}\xspace}

\newcommand{\sha}{{SHA-$256$}\xspace}
\newcommand\isve[1]{\texttt{Encl\textsubscript{#1}}\xspace}
\newcommand\pme[1]{\texttt{Encl\textsubscript{#1}}\xspace}
\newcommand\iron{{\sc Iron}\xspace}


\newcommand\sysname{{\sc Mage}\xspace}
\newcommand\premr[1]{\texttt{PREMR\textsubscript{#1}}\xspace}
\newcommand\mpg[1]{\texttt{MARS\textsubscript{#1}}\xspace}
\newcommand\mainfo[1]{\texttt{MAINFO\textsubscript{#1}}\xspace}
\newcommand\mpgsecname{{\tt .sgx\_mage}\xspace}
\newcommand\libname{{\tt libsgx\_mage}\xspace}

\newcommand{\gbytes}{\ensuremath{\mathrm{GB}}\xspace}
\newcommand{\mbytes}{\ensuremath{\mathrm{MB}}\xspace}
\newcommand{\kbytes}{\ensuremath{\mathrm{KB}}\xspace}
\newcommand{\bytes}{\ensuremath{\mathrm{B}}\xspace}
\newcommand{\hertz}{\ensuremath{\mathrm{Hz}}\xspace}
\newcommand{\ghertz}{\ensuremath{\mathrm{GHz}}\xspace}
\newcommand{\msecs}{\ensuremath{\mathrm{ms}}\xspace}
\newcommand{\usecs}{\ensuremath{\mathrm{\mu{}s}}\xspace}
\newcommand{\nsecs}{\ensuremath{\mathrm{ns}}\xspace}
\newcommand{\secs}{\ensuremath{\mathrm{s}}\xspace}
\newcommand{\gbits}{\ensuremath{\mathrm{Gb}}\xspace}

\newcommand\yz[1]{\textcolor{red}{\{\textbf{yinqian:} {\em#1}\}}}
\newcommand\gx[1]{\textcolor{green}{\{\textbf{guoxing:} {\em#1}\}}}
\newcommand\zq[1]{\textcolor{blue}{\{\textbf{Zhiqiang:} {\em#1}\}}}
\newcommand\scc[1]{\textcolor[rgb]{0.858, 0.188, 0.478}{\{\textbf{sanchuan:} {\em#1}\}}}

\newcounter{packednmbr}
\newenvironment{packedenumerate}{
\begin{list}{\thepackednmbr.}{\usecounter{packednmbr}
\setlength{\itemsep}{0pt}
\addtolength{\labelwidth}{4pt}
\setlength{\leftmargin}{12pt}
\setlength{\listparindent}{\parindent}
\setlength{\parsep}{3pt}
\setlength{\topsep}{3pt}}}{\end{list}}

\newenvironment{packeditemize}{
\begin{list}{$\bullet$}{
\setlength{\labelwidth}{8pt}
\setlength{\itemsep}{0pt}
\setlength{\leftmargin}{\labelwidth}
\addtolength{\leftmargin}{\labelsep}
\setlength{\parindent}{0pt}
\setlength{\listparindent}{\parindent}
\setlength{\parsep}{2pt}
\setlength{\topsep}{1pt}}}{\end{list}}

\newcommand{\cmark}{\ding{51}}%
\newcommand{\xmark}{\ding{55}}%

\title{\sysname: Mutual Attestation for a Group of Enclaves without Trusted Third Parties}

\author{\IEEEauthorblockN{Guoxing Chen,
Yinqian Zhang}
\IEEEauthorblockA{Department of Computer Science and Engineering\\
The Ohio State University\\
chen.4329@osu.edu, yinqian@cse.ohio-state.edu}
}


\maketitle

\begin{abstract}
Intel Software Guard Extensions (SGX) local and remote attestation mechanisms enable an enclave to attest its identity (\ie, the enclave measurement, which is the cryptographic hash of its initial code and data) to an enclave. 
To verify that the attested identity is trusted, one enclave usually includes the measurement of the enclave it trusts into its initial data in advance assuming no trusted third parties are available during runtime to provide this piece of information.
However, when mutual trust between these two enclaves is required, it is infeasible to simultaneously include into their own initial data the other's measurements respectively as any change to the initial data will change their measurements, making the previously included measurements invalid. 
In this paper, we propose \sysname, a framework enabling a group of enclaves to  mutually attest each other without trusted third parties. Particularly, we introduce a technique to instrument these enclaves so that each of them could derive the others' measurements using information solely from its own initial data. We also provide a prototype implementation based on Intel SGX SDK, to facilitate enclave developers to adopt this technique. 
\end{abstract}

\section{Introduction}

As storage and computation outsourcing to clouds become more and more prevalent, cautious users and security researchers raise questions on whether the cloud providers could keep their data private and execute their applications as expected. Trusted execution environments (TEEs) offer solutions to these concerns. A TEE is a secure component of a processor that protects the confidentiality and integrity of the code and data it executes upon. Examples of TEEs include Intel Software Guard Extensions (SGX), AMD Secure Encrypted Virtualization (SEV), and ARM TrustZone. Among them, Intel SGX, due to its convenient development and deployment model, is widely considered the most promising TEE solution, drawing attention from both industry and academia since its introduction~\cite{Anati:2013:sgxseal, McKeen:2013:sgxisolate, Hoekstra:2013:sgxsolution, Schuster:2015:vc3,Zhang:2016:towncrier, Tramer:2016:SGP, Ohrimenko:2016:OMP, Tamrakar:2017:CGS,Zheng:2017:opaque, Zhang:2017:rem, Baumann:2015:haven, Arnautov:2016:scone, Strackx:2016:ariadne, Hunt:2016:ryoan, Tsai:2017:graphene, Kuvaiskii:2017:SMS,Weiser:2017:sgxio,Tychalas:2017:sgxcrypter,Shinde:2017:Panoply,Seo:2017:sgxshield, Matetic:2017:ROTE}. Commercial clouds, such as Microsoft Azure~\cite{AzureSGX} and Alibaba Cloud~\cite{AlicloudECS} have offered SGX platforms for confidential cloud computing. A startup called Fortanix~\cite{fortanix} adopts SGX to provide runtime encryption solutions.


Intel SGX provides software applications a shield execution environment, called \textit{enclave}, to execution their proprietary code on secret data. This is achieved through a set of hardware and microcode extensions, including a special CPU execution mode (\ie, enclave mode),  an extended memory management unit that performs isolation during address translation, a memory encryption engine that sits between the CPU and the memory controllers. Intel SGX assumes privileged adversaries, such as malicious operating systems or rogue administrators. Therefore, the enclave memory is automatically encrypted with an encryption key known only to the CPU, and properly isolated so that it is not accessible to even the most privileged software. Only code inside the enclave region is able to access the enclave memory when executed under the enclave mode.

\bheading{Trusting an enclave via remote attestation.}
Before provisioning any secrets to an enclave, the enclave must be ``trusted''. This trust is established via remote attestation~\cite{inteliasapi}. In the context of attestation, the enclave is denoted the \textit{attester} and the entity that wishes to establish trust on the attester is denoted the \textit{attestee}, which could be either the user of the enclave (\ie, a human empowered by code) or \textit{another enclave}. Any software component between them can be considered as untrusted or even malicious. The establishment of trust can be achieved by answering the following three questions, the first two of which have been addressed by Intel SGX:

\begin{packeditemize}
\item \textit{Is the attester an enclave?} 
A particular private key, called \textit{attestation key}, is used to sign the message (a data structure called \textit{quote}) the attester sends to the attestee to prove that the producer of the message is indeed an enclave running on an SGX platform. The attestation key is endorsed by a root secret called \textit{root provisioning key}, which is burnt into the SGX processor during the manufacturing process, and could only be used to sign messages produced by enclaves. Hence, when the quote is verified to be valid, the attestee can be assured that the attester is indeed an SGX enclave. Intel SGX also provides another simpler mechanism, called \textit{local attestation}, to address the case when the attestee is also an enclave running on the same platform as the attester. 
\item \textit{What is its identity?} 
To address this question, Intel SGX adopts two types of identities: (1) the \textit{enclave identity}, \ie, the enclave measurement (\texttt{MRENCLAVE}), which is the cryptographic hash of the initial code and data of an enclave, and thus is used to identity the enclave's contents. The enclave measurement is calculated by the hardware when loading enclave code and data during enclave creation. Hence, the integrity of the enclave measurement is thus protected by the hardware; (2) the \textit{sealing identity} (\texttt{MRSIGNER}), which is the cryptographic hash of a public RSA key that identifies the enclave's developer.
Both identities are included in quotes. Hence, the attestee could obtain the attester's identities at the same time when verifying the quotes.
\item \textit{Is the identity trusted?} After the attestee is convinced that it is communicating with a real enclave with its specific identities \texttt{MRENCLAVE} and \texttt{MRSIGNER}, it is solely the attestee's decision whether an enclave with the given \texttt{MRENCLAVE} and/or \texttt{MRSIGNER} can be trusted. Note that a user of an enclave is not necessarily the developer of the enclave. For example, in the cases of running smart contracts or microservices in enclaves~\cite{Cheng:2019:Ekiden,Das:2019:FastKitten}, the users and the developers are different parties. If the user chooses to trust a developer, any enclave signed by the developer will be trusted. This is clearly not intended  in cases where the users and the developers are separate parties. In this paper, we consider a trust model where enclaves are trusted by their measurements (\texttt{MRENCLAVE}), not their developers (\texttt{MRSIGNER}).
Hence, the attestee needs to hold the measurement of a trusted attester in advance, so that the verification can be carried out easily. Particularly, when the attestee is also an enclave, the trusted attester's measurement will be hardcoded into its own initial enclave data~\cite{Alder:2019:sfaas, Fisch:2017:IRON, Chen:2019:OPERA}.
\end{packeditemize}

\bheading{Mutual attestation.}
Note that the aforementioned trust establishment is unidirectional, \ie, from an attester to an attestee. When the attestee is also an enclave, mutual attestation may be necessary. Mutual attestation is a mechanism that allows the communicating enclaves to attest each other and then establish a trust relationship. This is necessary, for example, in the following scenarios:

\begin{packeditemize}

\item Two enclaves from different developers running on the same machine authenticate each other. They trust less the reputation of the other developer, but the enclave identity of each other (after inspecting their code). 

\item An enclave running in a web server provisions secrets to an enclave running in the client's browser, while neither the web server nor the browser is trusted by the enclaves~\cite{Knauth:2018:sgxtls}. 

\item When data needs to be exchanged between SGX-enhanced privacy-preserving blockchains~\cite{Cecchetti:2017:Solidus, Das:2019:FastKitten, Cheng:2019:Ekiden, Lind:2019:Teechain}, an enclave from one blockchain needs to first attest the identity of another enclave from a different blockchain. 
\end{packeditemize}

\bheading{Mutual attestation without trusted third parties (TTP).} Mutual attestation can be achieved by simply performing attestation twice, one per each direction, by a trusted user. The user may also delegate this effort to a trusted third party (TTP). A TTP could be a stand-alone server that performs remote attestation with each enclave, validates the results (in collaboration with Intel's attestation services), and exchanges secrets with the two enclaves as a middle man to bootstrap the trust. However, integrating a user or a TTP into the application's operation dramatically increases the trusted computing base (TCB) of the entire application. The security of the application will hinge upon the trustworthiness of the software stack of the user or the TTP, rather than solely the security of the enclave code itself (and, of course, that of the CPU hardware). Therefore, it is often desired to perform mutual attestation without TTPs. 


In this paper, we aim to provide a mechanism for a group of (two or more) enclaves to mutually attest one another by their enclave identity. However, we found this problem non-trivial. Consider the cases of mutual attestation with two enclaves that would establish mutual trust with each other. The difficulty to do so lies in that
both enclaves need to wait for the other enclave's measurements to be finalized before they could include them in their initial data to finalize their measurements and release them. The situation has some similarity to the deadlock problem, in that both parties wait on the data held by each other before they can proceed. To our best knowledge, no prior work has addressed this problem.

\bheading{Our solution: \sysname.}
The key challenge of mutual attestation for a group of enclave without trusted third parties is to enable each of these enclaves to obtain the measurements of other enclaves in the same group from its own enclave memory, so that during the attestation phase, the enclave could verify whether the measurement of the attester is the same as one of the trusted enclaves in the same group. 
As such, we propose a framework, dubbed \sysname, to allow a group of enclaves to derive the measurements of other enclaves in the group from some \textit{intermediate states} instead of the final outputs of the other enclaves' measuring process.  
The key observation is that the measurement calculation is deterministic and sequential. Knowing intermediate states and information to perform subsequent measuring operations would be sufficient to derive the final output, \ie, the enclave measurement. When designed carefully, the problem could be resolved as all enclaves could generate intermediate states of their measuring processes and share them with others simultaneously.

Particularly, \sysname adds at the end of each enclave an extra data segment with the same content which includes the intermediate hash value of each trusted enclave's content right before the extra data segment. Hence, during runtime, each enclave knows the intermediate hash value of another trusted enclave (retrieved from the extra data segment) and the content left to be added (\ie, the extra data segment), and thus could derive the that trusted enclave's measurement. 
%
We have implemented a prototype of \sysname by extending the Intel SGX SDK. The evaluation suggests that to enable mutual attestation for up to $85$ enclaves, $62$ \kbytes enclave memory overhead for each enclave is introduced and roughly $21.7 \mu$s is needed to derive one measurement. We also plan to open-sourced it on Github. 
%
While the proposed scheme is originally designed for Intel SGX, the method can be easily extended to different types of TEEs, \eg, AMD SEV, and even between different types of TEEs, as long as they adopt similar mechanisms for the calculation of measurements.

\bheading{Paper outline.}
The rest of the paper is organized as follows: \secref{sec:background} provides necessary background for this paper. \secref{sec:motive_eg} presents motivating scenarios and \secref{sec:overview} gives an overview of the proposed scheme. The main component, the technique for enclaves to mutually derive each other's measurements is presented in \secref{sec:design}. We present a prototype implementation and evaluate the performance in \secref{sec:impl}. \secref{sec:discussion} discusses improvements and extensions. \secref{sec:related} presents related works and \secref{sec:conclusion} concludes this paper. 
\section{Background}
\label{sec:background}

\subsection{SGX Memory Organization}
Intel Software Guard Extensions (SGX) is a new hardware feature introduced on recent Intel processors. Intel SGX provides a shielded execution environment, called \textit{enclaves}, to protect sensitive code and data from untrusted system softwares. Particularly, Intel SGX reserves a specified range of DRAM, called \textit{Processor Reserved Memory} (PRM), which will deny accesses from any software (including the operating system) other than the enclave itself. 
The enclave's code, data, and related data structures are stored in a subset of PRM, called \textit{Enclave Page Cache} (EPC), which is further split into $4$ \kbytes EPC pages.


When creating an enclave, the SGX instruction \texttt{ECREATE} will be called to create the first EPC page, called \textit{SGX Enclave Control Structure} (SECS) page, which maintains the metadata of the enclave to be created, such as the base address, the size of enclave memory required, and the identity of the enclave. Then, via the  SGX instruction \texttt{EADD}, two types of EPC pages will be added: (1) \textit{Thread Control Structure} (TCS) pages that store information needed for logical processors to execute enclave code, such as the start address of enclave code when entering enclave mode via the SGX instruction \texttt{EENTER}; and (2) \textit{regular} (REG) pages that store the enclave's code, data and other related data structures such as \textit{State Save Area} (SSA), which is used to store enclave code's execution context during interrupts to protect the execution context from being learned by the untrusted system software (the processor will clear the execution context before transferring control to the interrupt handler and resume the execution context from the copy in the SSA after the interrupts are resolved and the enclave's execution is resumed). Note that when adding either a TCS page or a REG page, a $64$-byte data structure, called \textit{Security Information} (SECINFO), is also needed for the \texttt{EADD} instruction to specify the properties of the added EPC page, such as page type (a TCS page or a REG page), and access permissions (whether the page can be read, written and/or executed). 
After all enclave pages are loaded, the SGX instruction \texttt{EINIT} will be invoked to finalize the creation of the enclave. 
And the enclave code could be run then.


\subsection{SGX Enclave Measurements}
\label{sec:mrenclave}

\begin{figure}[t]
    \centering
\includegraphics[width=0.45\textwidth]{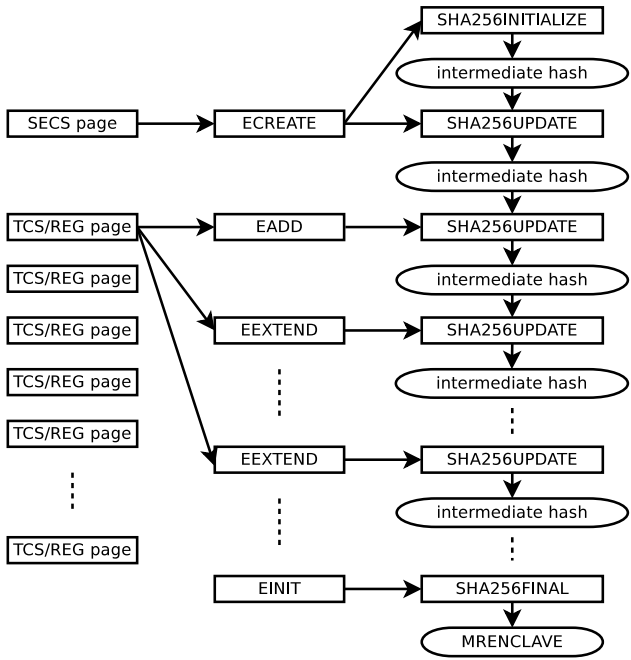}
	\caption{Enclave measurement calculation flow.}
	\label{fig:measurement}
\end{figure}

Intel SGX uses enclave measurements to identify enclaves. Generally, the enclave measurement is the cryptographic hash of the contents of an enclave, including initial code, data, and related data structures such as TCSs and SSAs. In this way, the enclave user could verify the identity of an enclave by checking only its measurement with an expected value.


In the current Intel SGX design, enclave measurements are calculated using \sha~\cite{Barker:SHA}. \sha is a Secure Hash Algorithm (SHA) that is used for generating $256$-bit digests of messages. The generated digests are used to protect the integrity of the messages. \sha has three algorithms: 
\begin{packeditemize}
\item \textit{Initialization algorithm} initialize $8$ $32$-bit words, called intermediate hash, before calculating the digest.
\item \textit{Update algorithm} takes a $512$-bit block as input at a time, and updates the intermediate hash using pre-defined compression functions.
\item \textit{Finalization algorithm} update the intermediate hash with the last $512$-bit block which contains the number of all bits that have been updated to the intermediate hash, and produce the final $256$-bit digest by concatenating the resulting $8$ $32$-bit words.
\end{packeditemize} 

\begin{table}
\caption{$512$-bit block updated to \texttt{MRENCLAVE} by \texttt{ECREATE}.}
\label{tab:ecreate}
\centering
\small
\begin{tabular}{p{.08\textwidth}|p{.28\textwidth}}
\Xhline{1pt}
\textbf{Range} & \textbf{Content} \\ 
\Xhline{1pt}
{[}$63:0${]}                           & $0045544145524345$H; // ``ECREATE''    \\ 
\Xhline{0.5pt}
{[}$95:64${]}                          & Size of one SSA frame in pages;                    \\ 
\Xhline{0.5pt}
{[}$159:96${]}                         & Size of enclave in bytes;                  \\ 
\Xhline{0.5pt}
{[}$511:160${]}                        & $0$;                                    \\ 
\Xhline{1pt}
\end{tabular}
\end{table}

Now we describe how Intel SGX leverages \sha to calculate enclave measurements. 
\texttt{MRENCLAVE} is a $256$-bit field located in the SECS page of an enclave. The calculation of \texttt{MRENCLAVE} is performed along with the creation of the enclave, as shown in \figref{fig:measurement}. When \texttt{ECREATE} is called to load an SECS page, the first page of an enclave, it also initializes the \texttt{MRENCLAVE} field using \sha Initialization algorithm and updates its value using \sha Update algorithm takes as input a $512$-bit block, including the metadata of the enclave such as the size of one SSA frame and the size of the enclave memory, as shown in \tabref{tab:ecreate}.

\begin{table}
\caption{$512$-bit block updated to \texttt{MRENCLAVE} by \texttt{EADD}.}
\label{tab:eadd}
\centering
\small
\begin{tabular}{p{.08\textwidth}|p{.28\textwidth}}
\Xhline{1pt}
\textbf{Range} & \textbf{Content} \\ 
\Xhline{1pt}
{[}$63:0${]}                           & $0000000044444145$H; // ``EADD''          \\ 
\Xhline{0.5pt}
{[}$127:64${]}                          & Offset of the added EPC page, relative to the enclave base;                     \\ 
\Xhline{0.5pt}
{[}$511:128${]}                        & The first $48$ bytes of SECINFO;   \\ 
\Xhline{1pt}
\end{tabular}
\end{table}

When \texttt{EADD} is called each time to create an TCS or REG page, it updates \texttt{MRENCLAVE} with a $512$-bit block as shown in \tabref{tab:eadd}. Note that \texttt{EADD} measures only the metadata of the page to be added, \eg, its offset and access permissions. The content of the page is measured by the SGX instruction \texttt{EEXTEND}. \texttt{EEXTEND} measures $256$ bytes at one time. As depicted in \tabref{tab:eextend}, For each $256$ bytes of an EPC page, \texttt{EEXTEND} performs the \sha Update algorithm $5$ times. The first iteration measures a $512$-bit block containing the metadata of the $256$ bytes  ($2048$ bits) of data including its offset, and the following $4$ iterations each  measures $512$ bits of these bytes. To measure a whole EPC page which consists of $4096$ bytes, $16$ \texttt{EEXTEND} operations are needed.

\begin{table}
\caption{Five $512$-bit blocks updated to \texttt{MRENCLAVE} for measuring the $256$-byte ($2048$-bit) data of an EPC page by \texttt{EEXTEND}.}
\label{tab:eextend}
\centering
\small
\begin{tabular}{p{.08\textwidth}|p{.28\textwidth}}
\Xhline{1pt}

\textbf{Range} & \textbf{Content} \\
\Xhline{1pt}
{[}$63:0${]}                           & $00444E4554584545$H; // ``EEXTEND''      \\ 
\Xhline{0.5pt}
{[}$127:64${]}                          & Offset of the $256$ bytes to be measured, relative to the enclave base;        \\ 
\Xhline{0.5pt}
{[}$511:128${]}                        & $0$;    \\ 
\Xhline{1pt}
\end{tabular}

\vspace{11pt}

\begin{tabular}{p{.08\textwidth}|p{.28\textwidth}}
\Xhline{1pt}
\textbf{Range} & \textbf{Content} \\ 
\Xhline{0.5pt}
{[}$511:0${]}                           & data[$511:0$];      \\ 
\Xhline{1pt}
\end{tabular}

\vspace{11pt}

\begin{tabular}{p{.08\textwidth}|p{.28\textwidth}}
\Xhline{0.5pt}
\textbf{Range} & \textbf{Content} \\ \hline
{[}$511:0${]}                           & data[$1023: 512$];     \\ 
\Xhline{1pt}
\end{tabular}

\vspace{11pt}

\begin{tabular}{p{.08\textwidth}|p{.28\textwidth}}
\Xhline{1pt}
\textbf{Range} & \textbf{Content} \\
\Xhline{0.5pt}
{[}$511:0${]}                           & data[$1535: 1024$];      \\ 
\Xhline{1pt}
\end{tabular}

\vspace{11pt}

\begin{tabular}{p{.08\textwidth}|p{.28\textwidth}}
\Xhline{1pt}
\textbf{Range} & \textbf{Content} \\ 
\Xhline{0.5pt}
{[}$511:0${]}                           & data[$2047: 1536$];      \\ 
\Xhline{1pt}
\end{tabular}

\end{table}

The last step of creating an enclave is to call the SGX instruction \texttt{EINIT} to finalize the measurement using the finalization algorithm of \sha. The finalization algorithm will update \texttt{MRENCLAVE} the last time with a $512$-bit block containing the total count of bits that have been updated into \texttt{MRENCLAVE}. This count is initialized during \texttt{ECREATE} and updated through \texttt{ECREATE}, \texttt{EADD} and \texttt{EEXTEND}.

\subsection{SGX Attestation}
For an enclave (attester) to authenticate its identity to another entity (attestee), Intel SGX provides two types of attestation mechanisms: local attestation and remote attestation. 
Generally, an attestation key will be used to generate a signature of the attester's identity along with a piece of attestation data. When the signature is verified to be valid by the attestee, it is assured that the attestation data is produced by the attester who is an enclave with the claimed identity running on an authentic SGX platform.

\subsubsection{Local Attestation}
Local attestation is introduced to enable an enclave to attest itself to another enclave located on the same platform. That is, both the attester and attestee are enclaves sharing the same processor. Specifically, the attester enclave calls the SGX instruction \texttt{EREPORT} to generate a \textit{Hash Message Authentication Code} (HMAC) signature of a data structure, called \textit{report}, which contains its own measurement and attestation data. 
Note that the calculation of HMAC involves a secret called \textit{report key}. Each report key is bond to an enclave, as an enclave measurement must be specified during the derivation of a report key.  
When calling \texttt{EREPORT}, the measurement of the target enclave (attestee) must also be provided. The processor will then derive the report key of the target enclave, to generate the HMAC. 
Note that this process is completely within the processor, and the report key of the target enclave is not exposed to the enclave memory of the attester during \texttt{EREPORT}. The only way to obtain the report key is to call the SGX instruction \texttt{EGETKEY} within the target enclave. \texttt{EGETKEY} will use the calling enclave's measurement to derive the report key. Hence, the report key is shared only between the target enclave and the SGX implementation.  
When a report is received and verified by the target enclave, it can be convinced that the report is generated by the SGX implementation on behalf of an attester whose measurement is also specified in the report.

\subsubsection{Remote Attestation}
To enable an enclave to attest itself to a remote entity, remote attestation is introduced. Currently, Intel adopts Enhanced Privacy ID (EPID) scheme for remote attestation. EPID is a digital signature algorithm that could protect the anonymity of SGX platforms~\cite{Intel:2016:EPID}. To facilitate EPID based remote attestation, Intel introduces two services, \ie, Intel Provisioning Service (IPS) and Intel Attestation Service (IAS), and provides SGX platforms with two privileged enclaves, Intel-signed Provisioning Enclave (PvE) and Quoting Enclave (QE). Particularly, Intel Provisioning Service and the Provisioning Enclave run an EPID provisioning protocol to provision an EPID private member key (attestation key) to an SGX platform. The EPID private member key could only be accessed by the Provisioning Enclave and the Quoting Enclave, otherwise, any malicious enclave that could access the EPID private member key will be able to forge valid signatures to convince the remote entity. Hence, to get a signature signed by the EPID private member key, the attester enclave needs to firstly attest itself to the Quoting Enclave via local attestation. After the Quoting Enclave verifies the attester's report, it will generate a data structure, called \textit{quote}, which contains the attester's measurement and attestation data that are copied from the report and sign the quote using the EPID private member key. The attestation enclave could then use the signed quote to attest itself to the remote entity. In the current design, the signed quote is encrypted by the Quoting Enclave, so that the remote entity has to forward the encrypted signed quote to Intel Attestation Service for verification. 

In December 2018, Intel introduced an Elliptic Curve Digital Signature Algorithm (ECDSA) based attestation solution~\cite{inteldcap}. Particularly, Intel provides a privileged enclave, called Provisioning Certification Enclave (PCE), which could access an ECDSA private key, called \textit{Provisioning Certification Key} (PCK), derived from within the SGX processor. Intel publishes the corresponding public key as an X.509 certificate for each SGX platform that supports ECDSA attestation. Hence, any party with the certificate could verify the ECDSA quote signed by Provisioning Certification Enclave run on the corresponding SGX platform.


\section{Motivating Scenarios}

\label{sec:motive_eg}

In this section, we present a few motivating scenarios, where mutual attestation of enclaves without trusted third parties is of great importance.

\subsection{Local Mutual Attestation with Enclave Identity}

Consider a case where two SGX enclaves running on the same machine need to establish mutual trust. Existing solutions are mainly based on the sealing identity (\texttt{MRSIGNER}) rather than the enclave identity (\texttt{MRENCLAVE}). Particularly, two mechanisms, \ie, local attestation and sealing, can be leveraged to achieve this:

\begin{packeditemize}

\item \textit{Trust via local attestation:}
As explained in \secref{sec:background}, local attestation can be used by one enclave to attest itself to another enclave. In the \textit{report} one enclave sends to another enclave via local attestation, both the enclave identity and sealing identity are included. Hence, the enclave that receives the report could compare only the sealing identity extracted from the report with the sealing identity of the trusted enclave, which should be hardcoded in its initial data in advance. Such local attestation could be run twice between the two enclaves, once per direction. Both enclaves need to have the other's sealing identity hardcoded in their initial data in advance. Note that this solution works for enclaves developed either by the same developer or different developers.
\item \textit{Trust via sealing:}
Intel SGX has provided a mechanism called \textit{sealing} for secret sharing between enclaves developed by the same developer that run on the same platform. Particularly, a seal key can be derived within the enclave to encrypt and decrypt secrets. During seal key derivation, the enclave could choose to include the identity of the developer, \ie, \texttt{MRSIGNER}, instead of \texttt{MRENCLAVE}. In this way, all enclaves signed by the same developer could derive the same seal key to decrypt the secret. The secret could be a shared private key used for establishing a secure channel. 
For SGX platforms that support key separation and sharing~\cite{IntelDevelopmentManual}, a configuration value and extra product IDs are introduced to further identifies enclaves developed by the same developer. 
Note that such sealing based solutions work only for enclaves developed by the same developer.
\end{packeditemize}

However, trusting the enclave developer could dramatically increase the trusted computing base (TCB). This is because any enclave built by the same developer shares the same sealing identity. An outdated, compromised enclave could be exploited to compromise other enclaves and hence the entire applications, if the trust between enclaves that established via their sealing identities. Finer-grain identification provided by key separation and sharing has the potential to reduce the attack surface, but its effectiveness is questionable as it highly relies on the developer's proper management of these configurations. 
On the other hand, when using the enclave identity during local attestation, only the enclave with the given measurement will be trusted.
Hence, exploring solutions that enable local mutual attestation with enclave identity could minimize the TCB that includes only the involved enclaves' code and data. 


\subsection{Remote Attestation for Server-Client Applications}
Intel SGX can be adopted in server-client applications to enhance the security of both the server and the client. In such scenarios, both the server and the client are equipped with enclaves that conceal sensitive data and code from the rest of the software stack, in order to minimize the TCB to only the SGX hardware and the enclave code. A secure channel established between the enclave on the client side and the enclave on the server side is desired to enable two-way authentication and secret provisioning.

For example, 
\texttt{OPERA} introduces an attestation service to provide better privacy guarantees to enclaves~\cite{Chen:2019:OPERA}. The proposed attestation service is based on EPID. It has two types of enclaves: issuing enclaves (\texttt{IssueE}) working as \textit{servers} that are responsible to provision EPID private keys to the other type of enclaves called attestation enclaves (\texttt{AttestE}), which function as \textit{clients}. \texttt{AttestE}{s} then use the provisioned EPID private key to provide attestation service to local enclaves. One important property of \texttt{OPERA} is its \textit{openness}, \ie, the implementation is completely open, so that its code (and hence behavior) can be publicly verified and thus is trustworthy, while its developer/signer (or sealing identity) can be untrusted. This property enables \texttt{OPERA} to achieve better security without introducing extra trusted parties. As such, an enclave identity based mutual attestation is desired in \texttt{OPERA}.

However, due to the lack of an enclave identity based mutual attestation mechanism between the server enclave and the client enclave, the authors of \texttt{OPERA} provided an alternative design that transfers part of the attestation workload to the user of the system. Particularly, only the server enclave verifies the identity (\ie the measurement) of the client enclave before provisioning EPID private keys. The client enclave has no means to verify whether the provisioned EPID private keys are from a trusted server enclave or not. The TCB of the server-client applications includes partial code on the user side that verifies whether the client enclave obtained the EPID private keys from a trusted server enclave or not.  While the authors proved the secrecy property of the protocol using \texttt{ProVerif}, they did not discuss other potential threats due to the lack of mutual attestation. For example, the adversary could provision the \texttt{AttestE} with an EPID private key controlled by the adversary and launch co-location attacks on the user of \texttt{OPERA} by monitoring the error message (\eg, ``EPID private key is from an untrusted server enclave'') of the attestation results. 
Hence, designing mutual attestation mechanism for server-client applications could reduce the attack surface and make the applications more self-contained. 

\subsection{Remote Attestation for Decentralized Applications}


Intel SGX has also been advocated as an enabling technology for privacy-preserving decentralized applications. The security of many decentralized applications, like Bitcoin's blockchain network and Tor's onion routing network, relies on the distributed trust over a large number of participating nodes, the majority of which are out of control of the adversary. SGX provides solutions to removing such trust completely from these participating nodes, protecting both the integrity and confidentiality of the secrets from these untrusted entities. For example,
Kim~\etal proposed SGX-Tor that utilizes SGX to enhance Tor~\cite{Kim:2017:SGXTor}. 
With SGX's strong confidentiality and integrity guarantees, SGX-Tor addresses some limitations of original Tor networks, \eg, weakening the threat model and mitigating low resource attacks. 

Moreover, designs that use SGX to provide privacy-preserving smart contracts have been proposed, such as Ekiden~\cite{Cheng:2019:Ekiden} and FastKitten~\cite{Das:2019:FastKitten}, with different focuses. Ekiden provides efficient off-chain execution of single-round contracts, while FastKitten targets efficient off-chain execution of reactive multi-round contracts. Hence, it is conceivable for one smart contract to delegate part of its execution to another smart contract that is more efficient in handling it. 
Establishing secure channels between the enclaves of these smart contracts requires mutual attestation and trusted third parties are undesired.
%
Furthermore, as blockchain interoperability, \ie, the ability to share data across different blockchain networks, is becoming more and more important for applications such as health care and voting, enabling these enclaves to mutually attest each other will be a fundamental problem to solve. Each enclave should be able the attest the other's identity before accepting the transaction executed by that enclave.

\section{Overview}
\label{sec:overview}

In this section, we describe the problem we aim to address, the threat model we assume, and the overall workflow of \sysname.

\subsection{Problem Formulation}

In this paper, we aim to address the problem of enabling a group of enclaves to mutually attest one another without trusted third parties. 
These enclaves could be developed by the same developer or multiple different developers, and 
the interaction between them are also specified by the code. These enclaves could be run on the same SGX platform (\ie, machine) or different SGX platforms; they need to mutually attest each other before they could start to interact and/or collaborate.


Intel SGX provides two types of attestation, \ie, local attestation and remote attestation. Both enable one enclave to verify that it is communicating with an actual enclave whose identity, \ie, the measurement, is in the received report (for local attestation) or quote (for remote attestation).  

Hence, the key challenge of mutual attestation is for each enclave to obtain the identities, \ie, measurements of the other trusted enclaves, without the help of trusted third parties. Consider a minimal group of two enclaves, \ie, \isve{1} and \isve{2}. For \isve{1} to verify an attester enclave is actually the other enclave \isve{2} in the group, \isve{1} has to know the measurement of \isve{2} in advance. Without a trusted third party to input \isve{2}'s measurement into \isve{1}, \isve{1} has to derive \isve{2}'s measurement by itself, \eg, by hard-coding \isve{2}'s measurement in its enclave memory. For mutual attestation, \isve{2} also needs to be able to derive \isve{1}'s measurement.

Simply hard-coding the other enclave's measurement in the enclave memory is not feasible. If we first hard-code enclave, say \isve{1}'s measurement into the other enclave \isve{2}'s initial data. Then we get the measurement of \isve{2} and try to hard-code it into \isve{1}'s initial data. However, this will change \isve{1}'s measurement. The previously hard-coded \isve{1}'s measurement in \isve{2}'s initial data will become incorrect. Observing that the measurement calculation is deterministic and sequential, we consider methods to derive the final measurement from the intermediate states and information required for the subsequent calculation. We now define the problem more formally as follows.

 Consider a group of $N$ mutually trusted enclaves, \isve{1}, \isve{2}, \dots, \isve{N}.
Denote the original content (code and data) of \isve{i} ($i = 1, \dots, N$) as $C_i$. The measurement of an enclave with content $C$ is the cryptographic hash of its content, denoted as $\mathcal{H}(C)$, where $\mathcal{H}$ is a cryptographic hash function.

\begin{definition}
\label{def:mdf}
A mechanism for mutual measurement derivation consists of two functions ($\mathcal{G}$, $\mathcal{F}$):
\begin{packeditemize}
\item $\mathcal{G}$ is called auxiliary content generation function that is used to generate auxiliary content needed for deriving measurements of other enclaves. It takes as input an index $i$ ($=1, \dots, N$) and the original contents of all $N$ enclaves, \ie, $C_1$, \dots, $C_N$, and output auxiliary content for \isve{i}, denoted as $A_i = \mathcal{G}(i, C_1, \dots, C_N)$. The content of \isve{i} then becomes the concatenation of $C_i$ and $A_i$, denoted as $C_i||A_i$. Its measurement becomes $\mathcal{H}(C_i||A_i)$.
\item $\mathcal{F}$ is called measurement derivation function that is used for deriving measurements from the auxiliary content. It takes as input the auxiliary content $A_i$ of \isve{i} ($i=1, \dots, N$) and an index $j$ ($=1, \dots, N$), and output the measurement of \isve{j}. Specifically, $\mathcal{F}$ satisfies \begin{equation}
\label{eq:def}
    \mathcal{F}(A_i, j) = \mathcal{H}(C_j||A_j), \forall i, j = 1, \dots, N
\end{equation}
\end{packeditemize}

\end{definition}

Note that in actual enclaves the layout of $C_i$ and $A_i$ might be interleaved instead of simple concatenation, we will discuss how to handle such situations in \secref{sec:discussion}.


\subsection{Threat Model}

We assume the hardware implementation of SGX is secure. That is, a malicious operating system cannot breach the confidentiality and integrity of the enclave code and data. At the time of writing, with proper microcode patches, known speculative execution attacks, such as Foreshadow~\cite{vanbulck2018foreshadow},  SgxPecture~\cite{SgxPectre} and Microarchitectural Data Sampling (MDS)~\cite{ridl,Schwarz2019ZombieLoad,fallout}, can no longer compromise the confidentiality of SGX enclaves. Moreover, Intel Attestation Service already provides the enclave users with information of SGX platform's microcode version (\ie, CPUSVN) and whether Hyper-Threading is disabled (for mitigating Foreshadow and MDS). Therefore, the attestation results could indicate whether the SGX platform is secure against speculative execution attacks.

We assume the code running inside of the enclaves is secure against memory corruption attacks~\cite{vanbulck:2019:taleoftwo,Biondo:2018:guarddilemma,Lee:2017:dark} 
and access-pattern-driven micro-architectural side channel attacks~\cite{Wang:2017:LCD, Schwarz:2017:MGE, Brasser:2017:SGE, hahnel:2017:HRS, Gotzfried:2017:CAI, Lee:2017:IFC, Van:2017:TYS,Xu:2015:CAD, Shinde:2015:PYF}. 
To securely use SGX, software programs must be thoroughly examined 
to be free of such vulnerabilities~\cite{wang:2019:timeandorder,xiao:2017:stacco}.


However, we assume SGX platforms are not trusted and may be controlled by the adversary. Specifically, the adversary controls all software components outside the enclave, including the operating system, the virtual machine manager (if any), the code running in the System Management Mode, \etc. The adversary is also able to launch any enclave as she wants; however, she cannot create an enclave whose measurement is pre-specified. Moreover, the adversary can perform man-in-the-middle attacks against the communication protocols between the enclaves, including but not limited to intercepting, dropping, replaying communications between any two enclaves.

This paper considers a group of trusted enclaves that would like to establish mutual trust between each other before communicating sensitive data among them. These enclaves are trusted by their measurements (\texttt{MRENCLAVE}) rather than their developers (\texttt{MRSIGNER}). The goal of the adversary is to gain the trust of the trusted enclaves via a malicious (enclave) program and earn the sensitive data. 


\subsection{Workflow of \sysname}

We now describe the workflow of \sysname, a framework enabling mutual attestation for a group of enclaves without trusted third parties, given a mechanism for mutual measurement derivation.

\begin{enumerate}
    \item Develop a system including a group of enclaves that need to interact or collaborate with each other;
    \item During compilation, derive auxiliary content from these enclaves and update each enclave with the derived auxiliary content;
    \item During runtime, enclaves derive the measurements to be used in either local attestation or remote attestation. 
\end{enumerate}

For the first step, a group of trusted enclaves that need to establish mutual trust are developed, especially when sensitive data needs to be transferred from one enclave to another. These enclaves could be developed by one or multiple developers. The algorithms about how the transferred secrets will be processed are up to the enclave developers, thus out of scope of this paper. The protocols for establishing secure channels include local attestation and/or remote attestation\cite{Shepherd:2017:MTC,Greveler:2012:MA}. \sysname will provide application programming interfaces (APIs) that implement a measurement derivation function $\mathcal{F}$ to be included in the attestation flows within the enclaves. 
The missing components are the auxiliary contents of other enclaves in the group to be used by the measurement derivation function to derive measurements of other enclaves.

Then, during compilation, a tool provided by \sysname, that implements the corresponding auxiliary content generation function $\mathcal{G}$ will be leveraged to extract auxiliary contents from these enclaves, and augment these enclaves with the derived auxiliary contents.  
After the augmentation, each enclave is ready to be signed by the developers and released.

Lastly, during runtime, whenever the measurement of a particular enclave in the group is needed, the measurement derivation API will be called to derive the measurement from the corresponding auxiliary content inserted earlier.

Note that most part of the workflow could fulfilled using existing SDKs and protocols. The missing and most critical component is the mechanism for mutual measurement derivation, \ie, the auxiliary content generation function $\mathcal{G}$ and the measurement derivation function $\mathcal{F}$ .
\section{\sysname Design}
\label{sec:design}

In this section, we introduce the mechanisms of \sysname. We will start with a minimal group of two enclaves and describe the necessary modifications needed for the enclave and the way to derive the other enclave's measurement. Then we discuss extension from two enclaves to a group of enclaves.

\subsection{Measurement Derivation from Intermediate Hash}

\begin{figure*}[t]
    \centering
\includegraphics[width=0.8\textwidth]{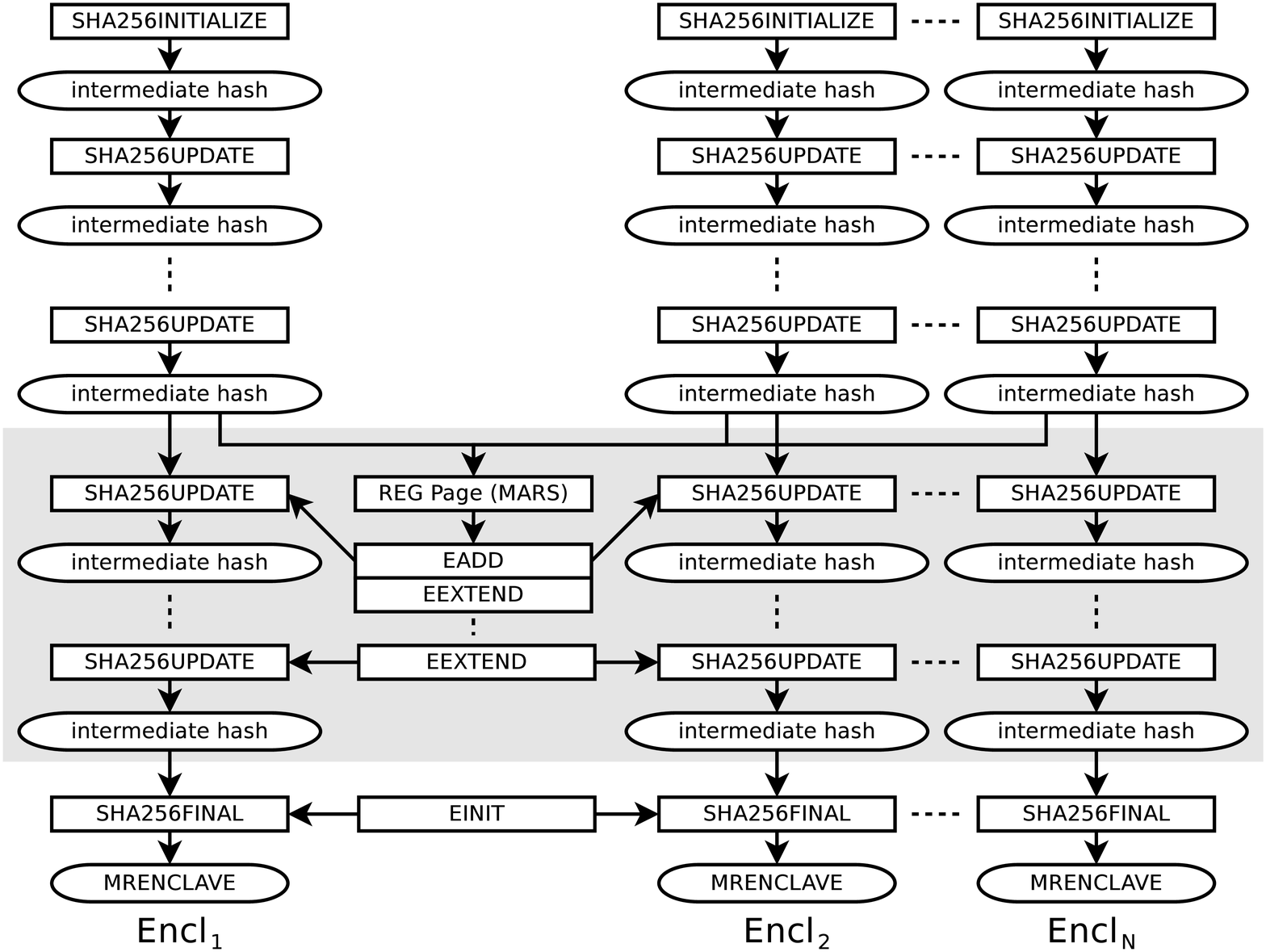}
	\caption{The flow of instrumenting the mutual attestation reserved segment \mpg{} with generated auxiliary content for measurement derivation.}
	\label{fig:design}
\end{figure*}

Now, we introduce another method for deriving measurements of enclaves with different contents. The key observation is that a  measurement, which is the cryptographic hash of an enclave's code and data, is calculated deterministically and sequentially. Setting aside the initialization and finalization phases, the hash of the concatenation of two messages $X$ and $Y$, denoted as $X||Y$,  is calculating the hash of $X$ first, and then calculating the hash of the concatenation of the resulting $\mathcal{H}(X)$ and $Y$. That is

\begin{equation}
    \mathcal{H}(X||Y) = \mathcal{H}(\mathcal{H}(X)||Y)
\end{equation}


When it comes to the enclave \isve{i} ($i=1, 2, \dots, N$) with content $C_i||A_i$. With any value set for the auxiliary content $A_i$, the calculation of the measurement satisfies the following equation:

\begin{equation}
    \mathcal{H}(C_i||A_i) = \mathcal{H}(\mathcal{H}(C_i)||A_i), \forall i = 1, 2, \dots, N
\end{equation}

Hence, we propose to include $\mathcal{H}(C_i)$ into the auxiliary content so that the measurement derivation could start from it. Specifically, we define an auxiliary content generation function $\mathcal{G}$ as

\begin{equation}
    \mathcal{G}(i, C_1, C_2, \dots, C_N) = \mathcal{H}(C_1)||\mathcal{H}(C_2)||\dots||\mathcal{H}(C_N)
\end{equation}

Note that the output $\mathcal{G}$ is irrelevant with regards to $i$, so the auxiliary contents of all these $N$ enclaves are the same, \ie, $A_i = A_j, i, j=1, 2, \dots, N$. Let $A_{i,j}$ represents the $j$-th hash value in $A_i$, which is $\mathcal{H}(C_j)$. We now define the measurement derivation function $\mathcal{F}$ as follows:
\begin{align*}
\label{eq:func}
    \mathcal{F}(A_i, j) &= \mathcal{H}(A_{i, j}||A_i) = \mathcal{H}(\mathcal{H}(C_j)||A_j)\\
    &= \mathcal{H}(C_j||A_j), \forall i, j = 1, \dots, N
\end{align*}



The above equation abstracted away details of actual \sha algorithms used in SGX. As described in \secref{sec:mrenclave}, for an actual enclave, given any intermediate hash value during the measurement calculation process, when the  information of all the remaining enclave pages to be added are provided, the enclave measurement can be derived by carrying out the subsequent \sha update and finalization operations from the intermediate hash value. For example, if both of \isve{1} and \isve{2} know each other's (1) intermediate hash before the last page to be added and the number of bits already updated to that intermediate hash; (2) the content, the offset and the SECINFO of the last page, they could derive each other's measurement.

Note that the content of the last page could be read (assuming that the access permissions allow read operations) by the enclave during runtime, we can calculate the intermediate hash of all pages before the last page in advance, and store the calculated value in the last page. In this way, both pieces of information are available during runtime. 
We will explain in more details in the following, starting from a group of two enclaves \isve{1} and \isve{2}.

\bheading{Remark.} There is no additional security risk to expose the intermediate hash values of these enclaves, as the intermediate hash values can be computed deterministically by any party who knows the enclave's initial code and data.

\begin{table}
\caption{\mainfo{}: information needed for measurement derivation.}
\label{tab:maise}
\centering
\small
\begin{tabular}{|p{.1\textwidth}|p{.3\textwidth}|}
\hline
\textbf{Component} & \textbf{Description} \\ \hline
\premr{}                           & The intermediate hash before \mpg{};          \\ \hline
\texttt{COUNT}                          &  The number of bytes updated to \premr{};                \\ \hline
\texttt{OFFSET}                     & The offset of \mpg{};   \\ \hline
\texttt{SECINFO}                     & The security information of \mpg{};   \\ \hline
\end{tabular}
\end{table}

\subsection{Compile-Time Auxiliary Content Derivation and Instrumentation}

Auxiliary content that is required for deriving the other enclave's measurement needs to be extracted and hardcoded into the enclave's initial data. It should be done during the enclave development phase. Basically, the \isve{1} and \isve{2} developers could follow the following steps:
\begin{packeditemize}
\item \isve{1} and \isve{2} developers reserve a data segment (called \textit{mutual attestation reserved segment}, denoted as \mpg{}) of the same size (\eg, $4$ \kbytes, the size of one EPC page) in the enclave memory, which will be loaded last during the enclave creation. Note that this data segment should be aligned to the page boundaries, so that it will not overlap with other enclave pages during measurement calculation. The \sha intermediate hash value of all pages before this reserved region will be calculated. We call this \sha intermediate hash value \textit{pre-measurement}, denoted as \premr{}.
\item \isve{1} and \isve{2} developers exchange information needed for derive their own measurements, called \textit{Mutual Attestation Information} (\mainfo{}) as depicted in \tabref{tab:maise}. Particular, \mainfo{} contains three fields: (1) the pre-measurement \premr{i} of \isve{i}, (2) the number of bytes updated to \premr{i}, (3) the offset of the reserved data segment \mpg{i}, and (4) the security information of \mpg{i}. The former two are used to reconstruct the state of measurement calculation before updating the reserved data segment \mpg{i}, and the latter two are needed for updating the \mpg{i} into the hash value as described in \secref{sec:mrenclave}. Note that \texttt{SECINFO} field can be dropped if the developers agree on the same SECINFO, \eg, assuming \mpg{} contains only read-only data pages. While fixing the offset of \mpg{} could also save the memory space for the \texttt{OFFSET} field, it will add extra workload for enclave developers to adjust the enclave memory layouts, which might not be preferred.
\item After the exchange, \isve{1} and \isve{2} developers organize the \mainfo{}{s} of both enclaves in the same order and instrument them into the \mpg{} of their own enclaves so that each enclave knows the \mainfo{} of the other enclave (from its own \mpg{}) and the content of the other enclave's \mpg{} (same as its own \mpg{}).
\end{packeditemize}

\subsection{Runtime Measurement Derivation} 

As described in \secref{sec:mrenclave}, the calculation of enclave measurements depends on the order the enclave pages are created. Even with exactly the same enclave code and data, when loaded in different orders, different measurements will be generated. Hence, during enclave creation, EPC pages of \isve{1} and \isve{2} have to be in a particular order that can be simulated during runtime to derive their measurements, as shown in \figref{fig:design}. Particularly, all enclave pages except \mpg{}{s} need to be created in the same order that generates the pre-measurement, \premr{}. The \mpg{}{} is created and loaded last. The enclave is initialized afterward. We will describe how to adjust the order of loading enclave pages in \secref{sec:impl} when needed, and also discuss an alternative design when the loading order cannot be altered. Now we assume the enclave is loaded exactly in the same order as how the \premr{} is computed.

\begin{figure}[t]
    \centering
\includegraphics[width=0.35\textwidth]{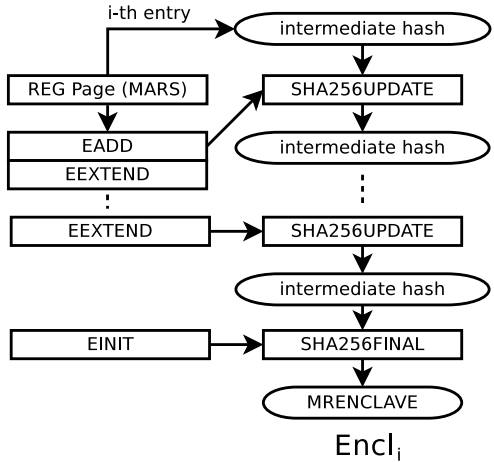}
	\caption{Runtime measurement derivation.}
	\label{fig:mdf}
\end{figure}

After one enclave, \eg, \isve{1}, is created, it could derive the measurement of the other enclave (\isve{2}) as shown in \figref{fig:mdf}:
\begin{packeditemize}
\item From \isve{1}'s reserved data segment \mpg{1}, \isve{1} retrieves \isve{2}'s pre-measurement \premr{2}, the number of bytes updated \premr{2}, the offset of \isve{2}'s reserved data segment \mpg{2}, and the SECINFO of \mpg{2}.
\item \isve{1} simulates \isve{2}'s process of loading the reserved data segment \mpg{2}. Note that the content of \mpg{2} is the same as \mpg{1} where \isve{1} could access directly. \isve{1} then update the number of bytes contributing to the resulting \sha intermediate hash and perform the finalization operation to obtain the measurement of \isve{2}.
\end{packeditemize}
For verification, let's recall how \isve{2}'s measurement is actually generated by the SGX implementation: \sha intermediate hash is updated as \isve{2}'s pages are created one by one; when it comes to \mpg{2} which will be loaded last, the \sha intermediate hash is \premr{2}, assuming the correct loading order; the \sha intermediate hash keeps being updated when loading \mpg{2} and gets finalized afterward. Hence, what \isve{1} derives is exactly the measurement of \isve{2}. Similarly, \isve{2} could also derive \isve{1}'s measurement.

\subsection{Supporting Multiple Enclaves}
Now we describe how to extend the method presented above to a group of (more than two) enclaves, \isve{1}, \isve{2},\dots, \isve{N}. Enclave developers need to extract and exchange the \mainfo{}{s} from their own enclaves. Then they organize the \mainfo{}{s} of all enclave in the same manner to generate identical \mpg{}{s}. And the creation of these enclaves needs to follow the pre-defined order for calculation \premr{}{s}. After one enclave is created, it could derive the measurement of any enclave by fetching the corresponding \mainfo{} from its own \mpg{} and simulating the measurement process with the content of its own \mpg{}. The measurement derivation function is shown in \algref{alg_1}. It takes as input the index $idx$ of the enclave measurement to be derived, and outputs the derived enclave measurement. The function retrieves the $idx$-th \mainfo{} to create an \sha handle, update it with the content of \mpg{} following the process described in \secref{sec:mrenclave}. Note that \texttt{sgx\_sha256\_init()}, \texttt{sgx\_sha256\_update()} and \texttt{sgx\_sha256\_get\_hash()} are the implementations of the initialization, update and finalization algorithms inside the enclave.

\begin{algorithm}[t]
\caption{Measurement Derivation Function}
\label{alg_1}
\KwIn{$idx$}
\KwOut{$mrenclave$}
\If{$idx$ $\ge$ total number of \mainfo{} entries in \mpg{}}{
    \Return NULL;
}
[\texttt{PREMR}, \texttt{COUNT}, \texttt{OFFSET}, \texttt{SECINFO}] $\gets$ $idx$-th \mainfo{} in \mpg{};\\
$sha\_handle$ $\gets$ sgx\_sha256\_init();\\
replace fields of $sha\_handle$ with \texttt{PREMR} and \texttt{COUNT};\\
\For{$page$ in \mpg{}}{
    sgx\_sha256\_update($sha\_handle$, ``EADD''$\|$\texttt{OFFSET}$\|$\texttt{SECINFO});\\
    \For {every $2048$-bit $data$ in $page$}{
        sgx\_sha256\_update($sha\_handle$,
        ``EEXTEND''$\|$\texttt{OFFSET});\\
        sgx\_sha256\_update($sha\_handle$,$data$[511:0]);\\
        sgx\_sha256\_update($sha\_handle$,$data$[1023:512]);\\
        sgx\_sha256\_update($sha\_handle$,$data$[1535:1024]);\\
        sgx\_sha256\_update($sha\_handle$,$data$[2047:1536]);
    }
    \texttt{OFFSET} = \texttt{OFFSET} + 256;
}
$mrenclave$ $\gets$ sgx\_sha256\_get\_hash($sha\_handle$);\\
\Return $mrenclave$;
\end{algorithm}

This solution is more scalable than the first one, as the overhead of adding one enclave is including only its \mainfo{} in \mpg{}. However, the scalability is still restricted by the maximum size of memory an enclave could have. To support even higher scalability requirements, we will discuss an alternative design when extra untrusted storage is available in \secref{sec:discussion}.


\subsection{Case Studies}

We now give a simple example of mutual attestation for migrating secrets from one enclave  (\pme{1}) to another enclave (\pme{2}). 
Considering secret migration could happen locally on the client's own computer with SGX support, we take local attestation as example to describe how to leverage \sysname to achieve secure secret migration. 

\begin{figure}[t]
    \centering
\includegraphics[width=0.45\textwidth]{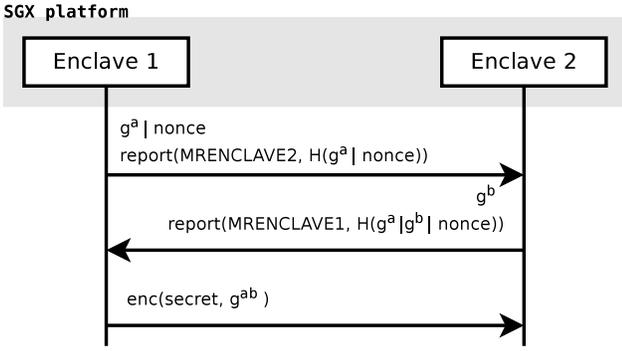}
	\caption{Establishing a secure channel for secret migration between two enclaves via local attestation.}
	\label{fig:pme}
\end{figure}

\figref{fig:pme} describes the workflow of establishing a secure channel (deriving a shared Diffie-Hellman
key) for secret migration between two enclaves via local attestation:

\begin{enumerate}
    \item After receiving a secret migration request from the client of both enclaves, \pme{1}     generates a Diffie-Hellman private/public key pair ($a, g^a$) and a nonce,  and sends the public key $g^a$ and the nonce to \pme{2} via local attestation. Particularly, \pme{1} derives the measurement of \pme{2} using the technique described in \secref{sec:design} and calls \texttt{EREPORT} to produce a report with the derived measurement as the target enclave measurement and the hash of $g^a$ and the nonce as the report data. $g^a$, the nonce and the report is sent to \pme{2}.
    \item After receiving the public key $g^a$, the nonce and the associated report, \pme{2} first verifies the identity of the sender by deriving the measurement of \pme{1} and comparing it with the one in the report structure, and then verifies the integrity of $g^a$ by calculating the hash of the received $g^a$ and the nonce, and comparing it with the hash in the report data. If verified valid, \pme{2} also generates a Diffie-Hellman private/public key pair ($b, g^b$), and sends the public key $g^b$ to \pme{1} via local attestation. Similarly, a report with the hash of both public keys and the nonce as report data and \pme{1}'s measurement as the target enclave measurement, along with the public key $g^b$ is sent to \pme{1}.
    \item Upon receiving $g^b$ and the corresponding report, \pme{1} verifies the identity of the sender and derives shared Diffie-Hellman key as $g^{ab}=(g^a)^b$. Hence, a secure channel between \pme{1} and \pme{2} is created. \pme{1} then encrypts the secrets to be migrated using $g^{ab}$ and sends the encrypted results to \pme{2}.
    \item \pme{2} could then decrypt the secrets. Optionally, \pme{2} might send back an acknowledgement message encrypted by $g^{ab}$, telling \pme{1} that the secrets are received and the copies on \pme{1}'s side could now be removed.
\end{enumerate}

\section{Implementation and Evaluation}
\label{sec:impl}
In this section, we describe our prototype implementation of \sysname and evaluate its runtime performance overhead and memory overhead.

\subsection{Implementation}
\sysname is implemented by extending the Intel SGX SDK (version $2.6.100.51363$)~\cite{sgxsdk}. Specifically, it consists of three components: (1) a SDK library that reserves a data segment for \mpg{} and APIs to support derivation of measurements from \mainfo{}{s} located in  \mpg{}; (2) a modified enclave loader that loads \mpg{}{s} last when creating enclaves; (3) modified signing tool that extracts \mainfo{} from an enclave and fills the \mpg{} of an enclave with a list of extracted \mainfo{}{s}.

\subsubsection{\sysname Library}
\libname is implemented to facilitate enclave developers to use \sysname. When included in an enclave, it reserves a read-only data section, named \mpgsecname, to be used as \mpg{}. The range of the \mpgsecname section is aligned to page boundaries, \ie, 4\kbytes. So its size is a multiple of the page size, \ie, 4\kbytes. Besides reserving the \mpgsecname section, \libname provides two APIs:
\begin{itemize}
\item \texttt{sgx\_mage\_size()} examines the \mpgsecname section and returns the total number of \mainfo{}{s} in it.
\item \texttt{sgx\_mage\_gen\_measurement()} takes as input an index of the enclave whose measurement is requested and outputs the resulting measurement. Particularly, it retrieves from the \mpgsecname section the corresponding \mainfo{} specified by the index and calculates the final measurement following \algref{alg_1}.
\end{itemize}

\subsubsection{Modified Enclave Loader}
The original enclave loader in Intel SGX SDK loads enclave code and data pages first and then the TCS pages. Hence, the \mpgsecname section, as a data segment, will not be loaded last by default. To address this, we modified the enclave loader to load the enclave pages in two stages:
\begin{itemize}
    \item First, the modified enclave loader follows the original loading process except that when an \mpgsecname section is encountered, it skips the \mpgsecname section. Note that \libname APIs are located in code pages and loaded along with the original enclave code pages.
    \item Second, when all other pages, including enclave code and data pages and the TCS pages, are loaded, the modified enclave loader checks whether there is an \mpgsecname section, and loads pages in the \mpgsecname section if found.
\end{itemize}

Note that if no \mpgsecname section is present, the modified enclave loader will load the enclave in the same order as the unmodified enclave loader, producing the same measurement. When there exists an \mpgsecname section, the modified and unmodified enclave loaders will produce different measurements due to the different loading order, as the unmodified enclave loaders will load \mpgsecname section earlier than the modified one. Since our implementation of \texttt{sgx\_mage\_gen\_measurement()} produces the measurement in the same order as the modified enclave loader, platforms running the enclaves developed with \sysname need to use to the modified enclave loader. If using modified loader is undesired, we provide an alternative design that works with unmodified enclave loaders. Details will be discussed in \secref{sec:discussion}.

\subsubsection{Modified Signing Tool}
The original signing tool is provided by Intel SGX SDK for enclave developers to sign enclaves, so that they can be accepted by the Intel-signed Launch Enclave and thus be launched successfully. The signing tool simulates the loading process of the enclave to calculate the measurement before signing it. We modified the signing tool to provide the following two functionalities:

\begin{itemize}
\item Deriving \mainfo{}{s}: given an enclave developed with \sysname, the modified signing tool could simulate the first stage of the modified enclave loader, which loads all pages except for the \mpgsecname section, to generate \mainfo{}, which includes the \sha intermediate hash, \ie, the \premr{}, the number of bytes updated to \premr{}, and the offset of the \mpgsecname section. The SECINFO is not included as our prototype implementation adopts a constant value of SECINFO with access permissions set to be read-only.
\item Filling the \mpgsecname section: given an enclave developed with \sysname and a set of \mainfo{}{s} derived from the group of trusted enclaves, the modified signing tool could fill the \mpgsecname section with the list of \mainfo{}{s}. The measurement of the instrumented enclave will be re-calculated and signed afterward.
\end{itemize}

\begin{figure}[t]
    \centering
\includegraphics[width=0.485\textwidth]{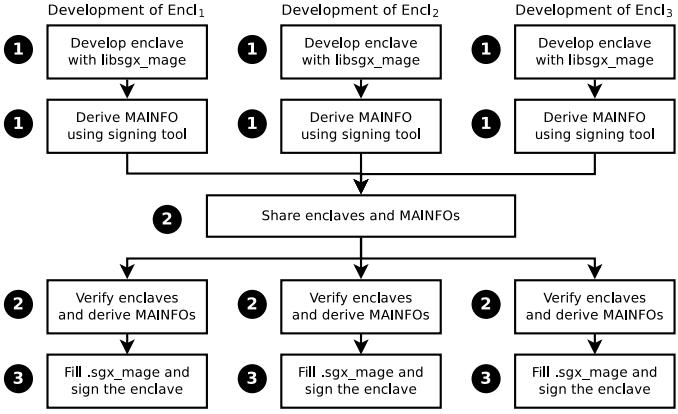}
	\caption{Workflow of enclave development using \sysname.}
	\label{fig:workflow}
\end{figure}

As such, the workflow of enclave development using \sysname (as shown in \figref{fig:workflow}) can be depicted as follows: \ding{202} the enclave developers independently implement their own enclave with the \libname library, and then use the modified signing tools to derive the \mainfo{} of their enclaves. \ding{203} the enclave developers share their enclaves with one another, so that they could validate the trustworthiness of the enclaves from other developers, and then use the modified signing tools to derive the \mainfo{}{s} of them. \ding{204} with the same list of \mainfo{}{s}, the enclave developers use the modified signing tool to fill the \mpgsecname section of their enclaves, and then sign their enclaves before publishing them. 

\subsection{Evaluation}

Now we describe the evaluation of our prototype implementation of \sysname. Results are measured on a Lenovo Thinkpad X1 Carbon ($4$-th Gen) laptop with an Intel Core i$5$-$6200$U processor and $8$GB memory. 


Since the results are highly related to the size of the \mpgsecname section, so we evaluate the metrics with regards to different sizes of the \mpgsecname section.

\subsubsection{The Number of \mainfo{}{s} Supported}
We first calculate the number of \mainfo{s} that can be stored in an \mpgsecname section with $L$ bytes ($L$ is a multiple of page size, \ie, 4\kbytes).
The content of an \mpgsecname section is organized as a structure as follows: the first $8$ bytes hold the total number of \mainfo{}{s} and the rest is used to store the content of these \mainfo{}{s}. Each \mainfo{} takes $48$ bytes ($32$-byte \premr{}, $8$-byte \texttt{COUNT}, and $8$-byte \texttt{OFFSET}). \texttt{SECINFO} is not included as we use a constant SECINFO in our prototype implementation. Hence, $\lfloor\frac{L-8}{48}\rfloor$ \mainfo{}{s} can be supported. For example, an \mpgsecname section of one page size could support up to $85$ \mainfo{}{s}. On the other hand, to support $N$ \mainfo{}{s}, a total of $\lceil\frac{48N+8}{4096}\rceil$ pages are needed. For example, supporting $N=10,000$ \mainfo{}{s} requires an \mpgsecname section of $=118$ pages ($472$ \kbytes). 

Note that the number of \mainfo{}{s} supported can be slightly increased when the ranges of \texttt{COUNT} and \texttt{OFFSET} are restricted. For example, SGX v1 has a limit of maximum enclave memory range for one enclave, which is $128$ \mbytes. Hence, any offset within the enclave can be represented using $27$ bits. Since the \mpgsecname section is aligned to page boundaries, $27-12=15$ bits are enough to store the offset of an \mpgsecname section. As for \texttt{COUNT}, loading one page could take up to $81$ \sha updates including $1$  \texttt{EADD} (consists of $1$ update) and $16$ \texttt{EEXTEND} (each includes $5$ updates). Hence, $15 + 7 = 22$ bits are enough for \texttt{COUNT}. $5$ bytes ($40$ bits) are enough to hold both \texttt{OFFSET} and \texttt{COUNT}. So one \mainfo{} can be as small as $32+5=37$ bytes so that an \mpgsecname section of one page size could hold $110$ \mainfo{}{s} instead of $85$.

\subsubsection{Efficiency of Measurement Derivation}

\begin{figure}[t]
    \centering
\includegraphics[width=0.485\textwidth]{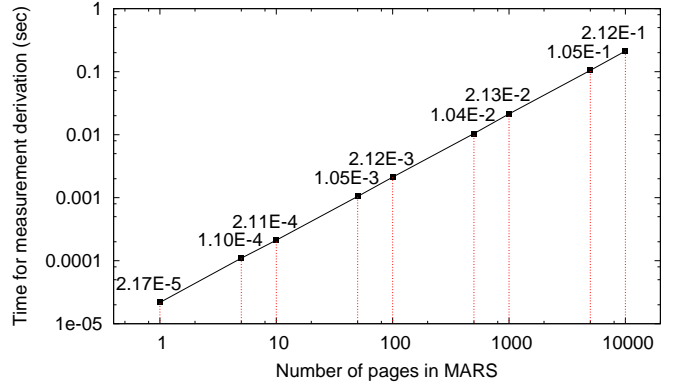}
	\caption{Measurement derivation efficiency.}
	\label{fig:perf}
\end{figure}

We then measure the time needed for deriving one measurement. From the measurement derivation function described in \algref{alg_1}, we can see that the time needed for the derivation is independent of the size of the original content of the enclave and the actual number of \mainfo{}{s} in the \mpgsecname section. This is because the content is updated into a single \mainfo{} where the derivation process starts from and all bytes in the \mpgsecname section need to be updated into the final measurement.

So we evaluated the efficiency of measurement derivation using a dummy enclave with only one enclave function that calls \texttt{sgx\_mage\_gen\_measurement()} to derive one measurement from its \mpgsecname section. Also, only one \mainfo{} from itself is generated and inserted into its \mpgsecname section. As expected, we verified that the derived measurement is the same as its own measurement. We measured the time (averaged from $10000$ iterations) needed to run one invocation of \texttt{sgx\_mage\_gen\_measurement()} when the number of pages in the \mpgsecname section ranges from $1$ to $10000$. The results are shown in \figref{fig:perf}. When the \mpgsecname section has a size of a single page, the time for deriving one measurement is around $2.17\texttt{e}{-5}$ seconds or $21.7 \mu$s. The time needed for deriving one measurement increases almost linearly with regards to the number of pages in the \mpgsecname section, because the main operations are updating the intermediate hash value using the content of \mpgsecname section.

\subsubsection{Memory Overhead}

\begin{table}[t]
  \caption{Memory overhead of \sysname}
  \label{tab:eval}
\centering
\small
\begin{tabular}{l|l|l|l|l|l}
\Xhline{1pt}
\# of pages in \mpgsecname                          & $1$                                    & $10$                                    & $100$                                    & $1000$                                    & $10000$                                   \\ 
\Xhline{1pt}
Size of \mpgsecname (\kbytes)                      & {\color[HTML]{000000} $4$}          & {\color[HTML]{000000} $40$}          & {\color[HTML]{000000} $400$}          & {\color[HTML]{000000} $4000$}          & {\color[HTML]{000000} $40000$}         \\ 
\Xhline{0.5pt}
Memory overhead (\kbytes)           & {\color[HTML]{000000} $62$}         & {\color[HTML]{000000} $98$}         & {\color[HTML]{000000} $458$}          & {\color[HTML]{000000} $4058$}          & {\color[HTML]{000000} $40058$}         \\ 
\Xhline{1pt}
\end{tabular}
\end{table}

The memory overhead introduced by \sysname includes two components: (1) extra data pages for \mpg{}; (2) extra code pages related to the measurement derivation. The first part is straight forward, which is the size of the \mpgsecname section. To calculate the second part, we created another enclave similar to the dummy enclave we just developed, except that the \sysname-related code and data are removed. We calculated the differences of memory sizes between these two enclaves. The results are presented in \tabref{tab:eval}. Subtracting the first component from the total extra memory, we got the size of the second component, which is around $58$ \kbytes. Since \libname leverages the \sha implementations provided in the Intel SGX SDK, the second component could be smaller if the original enclave already includes them.
\section{Discussion}
\label{sec:discussion}
In this section, we discuss possible extension of \sysname and its potential application to TEEs other than SGX.

\subsection{Extending \sysname with Untrusted Storage}

The basic design of \sysname enables the derivation of measurements from information completely inside the enclave memory. Next, we discuss how \sysname can be extended with untrusted storage outside the enclaves (\eg, unencrypted memory, hard drives, \etc.).  


\subsubsection{Supporting Unmodified Enclave Loaders}
In our basic design, we modified the enclave loaders to rearrange the order of enclave pages to be loaded in a way that the \mpg{} is loaded after all other contents. The \libname and the signing tool are modified to calculate measurements following the same order. While the \libname and the signing tool are used by enclave developers, the enclave loader runs on every SGX platform. Therefore, all parties must upgrade their toolchain to support \sysname. While we are discussing with Intel teams to merge \sysname into the official SDK, we next provide a temporary solution to support \sysname without the need of modifying enclave loaders and other SDK packages.

Consider the content of the enclave $C_i$ is split into two parts: $C_i^{pre}$ is the part loaded before $A_i$, and $C_i^{post}$ is the part loaded after $A_i$. The measurement can be calculated as
\begin{equation*}
    \mathcal{H}(C_i^{pre}||A_i||C_i^{post})
    =\mathcal{H}(\mathcal{H}(C_i^{pre})||A_i||C_i^{post})
\end{equation*}

We can still insert the intermediate hash of all pages before $A_i$, \ie, $\mathcal{H}(C_i^{pre})$ into $A_i$. As for $C_i^{post}$, instead of storing all its content within $A_i$, which introduces too much memory overhead, storing only the hash digest of $C_i^{post}$ within $A_i$ is more affordable. The auxiliary content generation function can be defined as 
\begin{align*}
    &\mathcal{G}(i, (C_1^{pre},C_1^{post}), \dots, (C_N^{pre},C_N^{post}))\\
    =& \mathcal{H}(C_1^{pre})||\mathcal{H}(C_1^{post})||\dots||\mathcal{H}(C_N^{pre})||\mathcal{H}(C_N^{post})
\end{align*}

During runtime, the enclave issues OCalls to request the host program to provide the content of $C_i^{post}$ for measurement derivation. The hash value $\mathcal{H}(C_i^{post})$ stored in $A_i$ will be used to verify the integrity of $C_i^{post}$ . Then, the measurement derivation function could start from $\mathcal{H}(C_i^{pre})$ and update the content of $A_i$ and $C_i^{post}$ into the intermediate hash to obtain the final measurement. This design requires extra untrusted storage to store $C_i^{post}$, which unfortunately may take longer time to derive a measurement when $C_i^{post}$ is large.

\subsubsection{Increasing Scalability}

\begin{figure}[t]
    \centering
\includegraphics[width=0.45\textwidth]{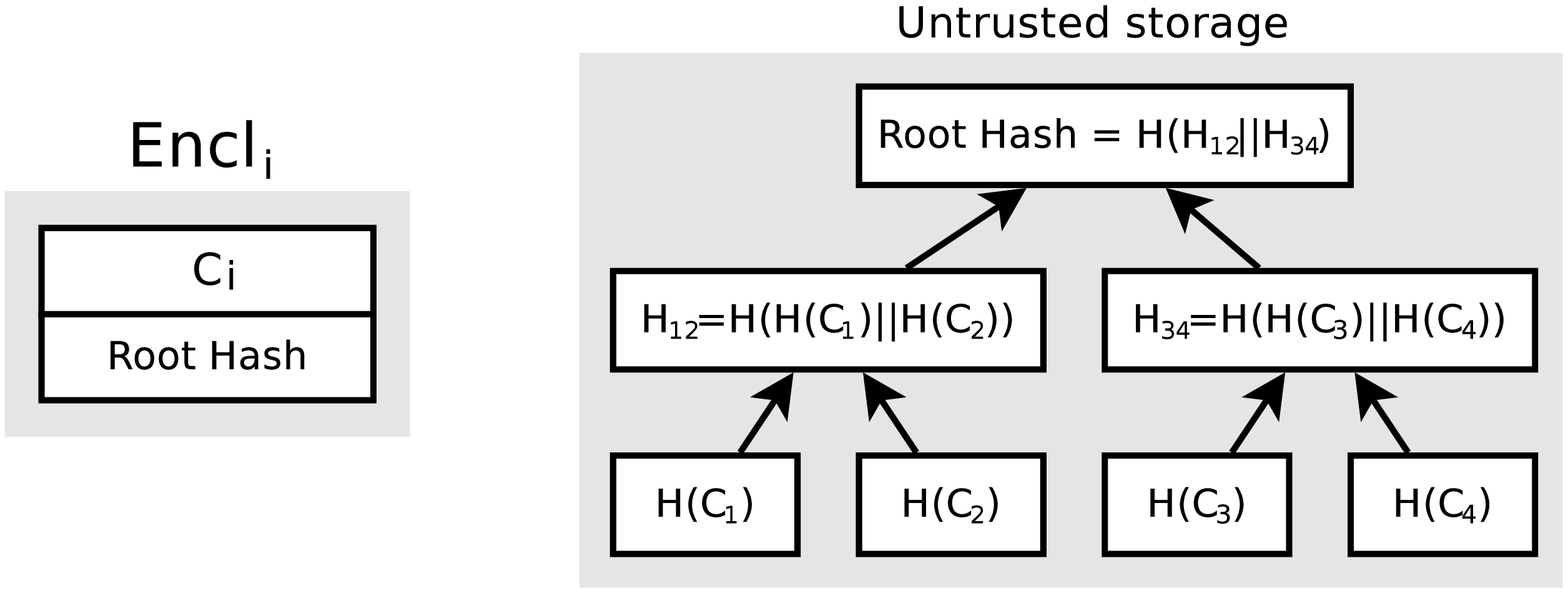}
	\caption{Alternative design for better scalability.}
	\label{fig:scalability}
\end{figure}

In our basic design, the time cost for measurement derivation and memory overhead grows linearly with regards to the size of the group of trusted enclaves. This would be a problem when the group becomes much larger. Especially, SGX v1 has a limit on the maximum size of enclave memory. To support a large number of enclaves, when untrusted storage outside the enclave is available, all \mainfo{}{s} can be moved out of the enclave memory and only the hash of all these \mainfo{}{s} is stored within the \mpg{} for integrity protection. During measurement derivation, the target \mainfo{} will be retrieved from the untrusted storage and authenticated within the enclave. In this way, the memory overhead becomes constant, as only one page of \mpg{} is needed to hold the hash value of all \mainfo{}{s}. The time cost for measurement derivation will also become constant as only one page of \mpg{} is required to be updated to the measurement.

However, since \mainfo{}{s} are stored in untrusted storage, the overhead for the \mainfo{} retrieval and integrity verification might still have a linear time complexity when sequential hashing algorithm such as \sha is used. To address this, as shown in \figref{fig:scalability}, the Merkle tree structure could be adopted to organize \mainfo{}{s} outside then enclave memory (only root hash of the Merkle tree is stored within the \mpg{}) for efficient retrieval and verification, achieving a logarithmic time complexity instead of a linear time complexity~\cite{Merkle:1987:MerkleTree}.


\subsection{Extensions to Other TEEs}

\sysname can be extended to other TEEs that use hash-based measurement mechanisms for attestation. For example, AMD's Secure Encrypted Virtualization (SEV) is a TEE solution that encrypts the memory of virtual machines (VM) without a trusted hypervisor. SEV also uses \sha digest of the guest memory to compute the measurements of guest VMs~\cite{AMDSEV}. 
Moreover, Sanctum is an open source RISC-V based TEE solution that offers similar promises as Intel SGX, which also adopts a measurement mechanism similar to SGX~\cite{Costan:2016:sanctum}. 
These TEEs could be integrated with \sysname to enable mutual attestation without TTP. While ARM TrustZone does not provide integrity measurement inherently, Zhao~\etal proposed a software-based to provide secure enclaves using TEE such as ARM TrustZone~\cite{Zhao:2019:SecTEE}. The proposed scheme also includes a hash-based measurement. Hence, \sysname can also be applied.

\begin{figure}[t]
    \centering
\includegraphics[width=0.45\textwidth]{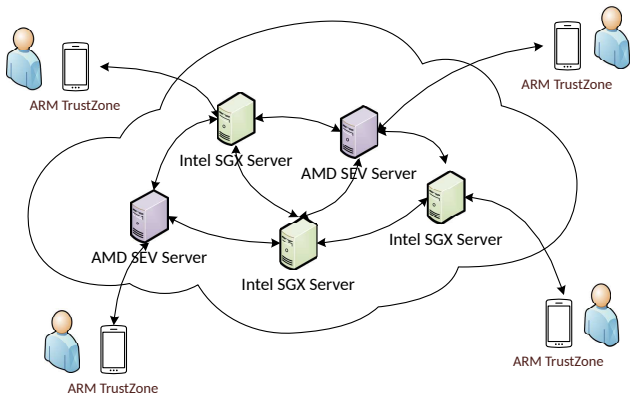}
	\caption{Privacy-Preserving Pandemic Tracking System.}
	\label{fig:ppml}
\end{figure}

Further, \sysname can be extended for different types of TEEs to mutually attest each other. This could benefit applications that integrate different types of TEEs. For example, as shown in \figref{fig:ppml}, a privacy-preserving pandemic tracking system could be possible when mobile devices with ARM TrustZone are used to collect and transmit users' trajectories to cloud platforms with Intel SGX via a secure channel established through mutual attestation. The collected trajectories could be monitored and analyzed privately within enclaves, and notifications would be returned to those affected mobile users. We expect mutual attestation between SGX and Trustzone would enable many other interesting use cases with cloud/client and edge/client computing models.


\subsection{Supporting Enclave Updates}

Although it is possible for the enclave code to be updated after they are deployed, for reasons like fixing bugs or introducing new functionalities, \sysname does not support enclave updates. If the content of any of the enclaves is changed, to continue the use of \sysname, all other enclave binaries need to be updated to reflect the change, before these enclaves are re-deployed in the system.

The lack of support of enclave updates in \sysname is intended. This is because updates of enclave code change not only the identity of the enclave (\texttt{MRENCLAVE}) but also the trustworthiness of its behavior, especially when the developers are not trusted, which is assumed in our model. Therefore, a new version of an enclave should be inspected again for its trustworthiness. In other words, the trust relationship between these enclaves should be re-evaluated if one has been updated. %

We note that enclave applications are similar to dApps built atop smart contracts, which are also difficult to patch once deployed. Therefore, solutions for dApps might also work for enclave applications. We leave the investigation of facilitating enclave updates to future work.

\section{Related Work}
\label{sec:related}



The work that is most related to ours is presented by Greveler~\etal~\cite{Greveler:2012:MA}. The authors proposed a protocol for two identical Trusted Platform Modules (TPMs) to mutually attest each other for system cloning. The two identical TPMs generate the same value of platform configuration register (PCR), which is cryptographic hash of the software loaded into the TPM, having the same functionality as the measurement in Intel SGX. 
However, this protocol only works when both entities have the same identity, \eg, PCR or measurement, so that each entity could simply use its own measurement for verification.  In contrast, our scheme enables enclaves with different measurements to mutually attest each other, enabling applications beyond system cloning.
Shepherd~\etal proposed a Bi-directional Trust Protocol (BTP) for establishing mutually trusted channels between two TEEs ~\cite{Shepherd:2017:MTC}. But BTP assumed that both TEEs know the identity of the other, while our work answers how this assumption could be realized.

Apache Teaclave, an open source universal secure computing platform, address the mutual attestation problem by relying on third-party auditors~\cite{teaclave:ma}. Enclaves will be audited and then signed by these auditors. The public keys of the auditors are hardcoded in these enclaves to support mutual attestation. These auditors are trusted by all involved enclaves and act as trusted third parties. On the contrary, \sysname tackles the mutual attestation problem without trusted third parties.

Also related to our work is a line of research on enclave migration. Park~\etal was the first to address this problem by proposing a new SGX hardware instruction to be used to produce a live migration key between two SGX platforms for secure transfer of enclave content~\cite{Park:2016:LiveMigration}. Gu~\etal proposed a software-based solution by augmenting enclaves with  a thread that could run remote attestation to establish a secure channel with the thread within another identical enclave, and then perform state transfer~\cite{Gu:2017:LiveMigration}. Alder~\etal proposed an approach to migrate the persistent states of enclaves, \eg, sealed data, which is outside of the enclave memory~\cite{Alder:2018:MigrSGX}. And Soriente~\etal designed ReplicaTEE for seamless replication of enclaves in clouds~\cite{Soriente:2019:ReplicaTEE}. While all these designs address secret migration between enclaves with the same measurement, our technique could complement them by enabling secret migration between  enclaves 
with different measurements.





\section{Conclusion}
\label{sec:conclusion}
In this paper, we study techniques for a group of enclaves to mutually attest each other without trusted third parties. The main contribution of this paper is the mutual measurement derivation mechanisms, enabling enclaves to derive other trusted enclaves' measurements during runtime. We implement the proposed mechanisms based on Intel SGX SDK and evaluate the performance. We demonstrate through case studies that this technique could facilitate new applications that require mutual trust for interaction and collaboration.

\balance
\bibliographystyle{IEEEtran}
\bibliography{paper}

\balance

\end{document}